\pdfoutput=1
\documentclass[11pt]{article}

\usepackage[T1]{fontenc}
\usepackage[utf8]{inputenc}
\usepackage[a4paper,margin=1in]{geometry}
\usepackage{amsmath,amssymb}
\usepackage{graphicx}
\usepackage{booktabs}
\usepackage{float}
\usepackage{hyperref}
\graphicspath{{figures/}}

\newcommand{\ket}[1]{|#1\rangle}
\newcommand{\bra}[1]{\langle #1|}
\newcommand{\D}{\mathcal{D}}
\newcommand{\Tr}{\operatorname{Tr}}

\title{Finite-Bandwidth Protection of a Three-Level Quantum Heat Engine Against Parasitic Heat Leaks}

\author{
Gilberto Aparecido Prataviera$^{1}$ and Marcos C\'esar de Oliveira$^{2,*}$\\[0.6em]
\small $^{1}$Departamento de Administra\c{c}\~ao, Faculdade de Economia, Administra\c{c}\~ao e Contabilidade\\
\small de Ribeir\~ao Preto (FEA-RP), Universidade de S\~ao Paulo, 14040-905 Ribeir\~ao Preto, S\~ao Paulo, Brazil\\
\small \texttt{prataviera@usp.br}\\
\small $^{2}$Instituto de F\'isica Gleb Wataghin, Universidade Estadual de Campinas,\\
\small 13083-859 Campinas, S\~ao Paulo, Brazil\\
\small $^{*}$Correspondence: \texttt{marcos@ifi.unicamp.br}
}

\date{}

\begin{document}

\maketitle

\begin{abstract}
Finite-bandwidth reservoir engineering can suppress unwanted transitions in a
quantum thermal machine, but a physical filter also introduces a finite response
time. We study this competition in a continuous three-level heat engine whose
hot environment couples parasitically to the cold transition. The corresponding
three-state rate network is solved exactly, showing that the parasitic transition
produces a hot-to-cold thermodynamic short circuit and yielding a closed
threshold for the loss of positive-power operation. We then retain a damped
auxiliary mode explicitly as a physical spectral filter. Its Lorentzian response
suppresses the detuned parasitic transition, whereas excessive narrowing limits
the useful energy throughput. Independent local-GKSL and nonsecular
Bloch--Redfield calculations both recover the no-leak Markovian engine in the
weak-coupling limit and predict a finite maximum-power bandwidth, although its
precise location is model dependent. A frequency-resolved Lorentzian-rate
reduction, by contrast, has no interior optimum. The resulting design principle
is that spectral selectivity can protect the useful thermodynamic cycle, but a
real filter cannot be narrowed without a dynamical throughput cost.
\end{abstract}


\section{Introduction}

The three-level maser is a canonical model of a continuous quantum heat engine
~\cite{Scovil1959,Alicki1979}.  Its thermodynamic cycle can be expressed directly
in terms of three transition frequencies.  A~hot reservoir excites the system
through a transition of frequency {$\omega_h$},
 a~cold reservoir exchanges energy
across a transition of frequency $\omega_c$, and~the remaining transition,
\begin{equation}
\epsilon=\omega_h-\omega_c,
\label{eq:epsilon}
\end{equation}
is connected to a work terminal.  In~the ideal tightly coupled model, every
stationary cycle absorbs $\omega_h$ from the hot reservoir, releases $\omega_c$
to the cold reservoir, and~transfers $\epsilon$ through the work terminal.  The~corresponding ideal efficiency is
\begin{equation}
\eta_0=\frac{\epsilon}{\omega_h}.
\label{eq:ideal_eff}
\end{equation}
Broader accounts of the theoretical and experimental developments in quantum heat engines and refrigerators can be found in Refs. \cite {heat1, heat2,heat3} and references~therein.

\begin{figure}[H]
\centering

\centering
\includegraphics[width=\textwidth]{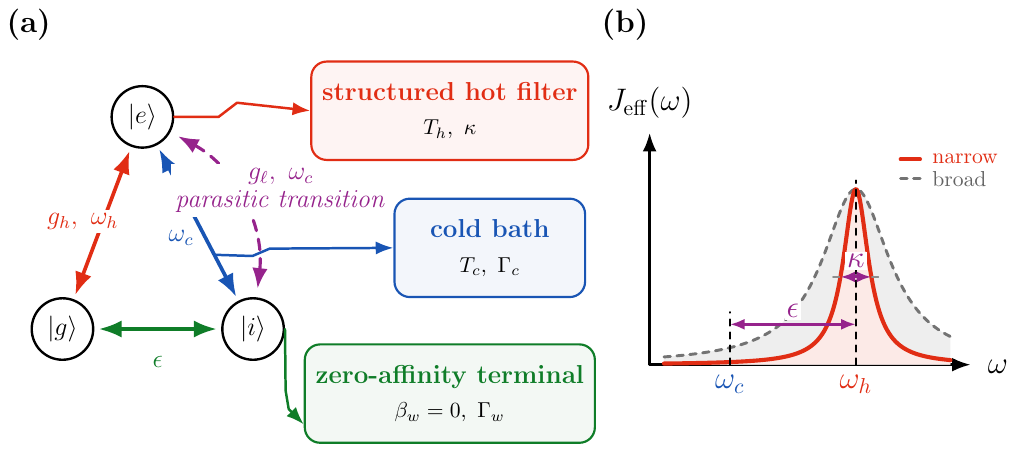}
\caption{{Engine}
 cycle and finite-bandwidth protection.  (\textbf{a}) The useful hot
transition $\ket g\leftrightarrow\ket e$ is resonantly coupled to a structured
hot filter, while the dashed purple line denotes the unwanted coupling of the
same hot filter to the cold transition $\ket i\leftrightarrow\ket e$.  The~cold
bath and the zero-affinity terminal act on the remaining two transitions.  (\textbf{b}) Narrow
and broad Lorentzian responses centered at $\omega_h$, shown at equal resonant
height as imposed by $4g_{h}^2/\kappa=\Gamma_h$.  Narrowing the filter suppresses
the detuned response at $\omega_c=\omega_h-\epsilon$, while an excessively
narrow filter becomes dynamically slow.}
\label{fig:engine_cycle}
\end{figure}

This picture assumes perfect transition selectivity.  However, in~an actual physical device, an~environment designed to address one transition generally has a finite spectral
width and can also drive a nearby transition.  In~the three-level engine
considered here, the~hot environment couples parasitically to the transition
normally addressed by the cold bath (see Figure~\ref{fig:engine_cycle}).  The~additional transition opens a second
thermodynamic cycle that transfers heat from hot to cold without crossing the
work transition, thereby breaking tight coupling and potentially reversing the
useful~cycle.

A natural strategy is to engineer a frequency-selective hot environment.
Reaction-coordinate, pseudomode, and~auxiliary resonator constructions provide
physical realizations of structured reservoirs
~\cite{Strasberg2016,Restrepo2018,AntoSztrikacs2021,Cavaliere2022}.  Related
filtering ideas have been explored in quantum refrigerators and thermal
management~\cite{He2017,Naseem2020}, and~superconducting thermal circuits offer
an experimentally relevant setting in which resonators shape the spectrum seen
by a small quantum system~\cite{Rasola2025,Uusnakki2026}.

The central issue is that spectral selectivity and dynamical response are not
independent when the filter is a physical subsystem. A~broad filter exchanges
energy rapidly with its reservoir but poorly resolves the useful and parasitic
transition frequencies. A~narrow filter provides stronger spectral rejection,
but its longer response time can itself limit the useful heat current. The~competition between these two effects suggests that the best operating point
may occur at a finite filter~linewidth.

In this work, we establish how finite-bandwidth reservoir engineering can
protect a continuous three-level quantum heat engine against a parasitic hot
transition while accounting explicitly for the dynamical cost of the filter.
We first formulate the three-level engine as an exactly solvable Markovian
rate network including the parasitic hot transition. This allows us to prove
that the parasitic transition generates a hot-to-cold thermodynamic short
circuit and to derive a closed threshold for the loss of positive-power
operation. We then replace the idealized frequency-dependent hot reservoir
by an explicit damped auxiliary mode that acts as a physical spectral filter.
Its finite linewidth suppresses the off-resonant parasitic transition, while
its finite response time limits the useful energy throughput when the filter
becomes too~narrow.

The quantitative behavior of the resulting qutrit--filter system can depend
on how its residual environments are modeled. We therefore compare several
open-system descriptions. Our primary calculation uses a local
Gorini--Kossakowski--Sudarshan--Lindblad (GKSL) master equation
~\cite{Gorini1976,Lindblad1976} for the enlarged system. As~an independent
robustness check, we also employ a frequency-resolved nonsecular
Bloch--Redfield treatment in the dressed basis, retaining interference between
near-degenerate transitions
~\cite{Redfield1957,Farina2019,Trushechkin2021}. Both explicit filter
descriptions are compared with a frequency-resolved Lorentzian-rate reduction
obtained from the filter susceptibility. The~latter can be connected
analytically to the three-state model and, unlike the explicit dynamical
descriptions, is shown not to possess an interior maximum of the stationary
power. We additionally consider a fully secular global construction only as a
diagnostic of the near-degenerate weak-coupling limit.
The two explicit filter approaches independently predict a finite
maximum-power bandwidth, although~its precise location is
model dependent. We further examine how this optimum changes with the
parasitic coupling strength and thermal bias, and~discuss a possible
superconducting-circuit implementation using a strongly anharmonic artificial
atom coupled to a tunable low-$Q$ resonator.

The remainder of the paper is organized as follows.
{Section}
 \ref{sec:markov} solves the Markovian multicycle engine and
derives the heat leak theorem and the engine breakdown threshold.
Section~\ref{sec:filter} introduces the explicit finite-bandwidth
filter and derives its time-domain response and Lorentzian spectrum.
Section~\ref{sec:methods} presents the local-GKSL,
nonsecular Redfield, and~effective-rate descriptions on a common
footing. Section~\ref{sec:validation} gives the numerical construction
and the weak-coupling and Fock-space validation tests.
Section~\ref{sec:results} compares the bandwidth dependence and
parameter trends predicted by the different methods.
Section~\ref{sec:discussion} discusses what is robust and what remains
model dependent, and~Section~\ref{sec:conclusion} summarizes the
conclusions.

\section{Three-Level Engine with a Parasitic Hot~Transition}\label{sec:markov}

We begin by defining the three-level working medium and the transition rates that will be used throughout the analysis.
We work in units in which $\hbar=k_B=1$.  The    three energy eigenstates are
denoted by $\ket g$, $\ket i$, and~$\ket e$, with~\begin{equation}
E_g=0,\qquad E_i=\epsilon,\qquad E_e=\omega_h,
\label{eq:levels}
\end{equation}
so that the energy gap between $\ket i$ and $\ket e$ is
\begin{equation}
\omega_c=\omega_h-\epsilon.
\label{eq:omega_c}
\end{equation}
The useful hot transition is $\ket g\leftrightarrow\ket e$, the~cold
transition is $\ket i\leftrightarrow\ket e$, and~the work transition is
$\ket g\leftrightarrow\ket i$.  In~addition, the~hot reservoir acts
parasitically on the $\ket i\leftrightarrow\ket e$ transition.  The~geometry
and spectral-filtering mechanism are summarized in Figure~\ref{fig:engine_cycle}.

Let $a$ and $b$ denote, respectively, the~hot excitation and relaxation rates
on $\ket g\leftrightarrow\ket e$.  Let $c_c$ and $d_c$ denote the cold
excitation and relaxation rates on $\ket i\leftrightarrow\ket e$, and~let
$c_\ell$ and $d_\ell$ denote the additional excitation and relaxation rates
produced by the hot reservoir on the same transition.  The~work terminal is
represented by equal rates $w$ in the two
directions, corresponding to an unbiased (zero-affinity) channel.
This channel is used operationally to define the stationary output
power. It is
not intended as a microscopic model of a battery or of ergotropic work
storage.

The state population equations are given by
\begin{align}
\dot p_g&=-a p_g+b p_e+w(p_i-p_g),
\label{eq:pgdot}\\
\dot p_i&=(d_c+d_\ell)p_e-(c_c+c_\ell)p_i+w(p_g-p_i),
\label{eq:pidot}\\
\dot p_e&=a p_g+(c_c+c_\ell)p_i-(b+d_c+d_\ell)p_e.
\label{eq:pedot}
\end{align}
Probability conservation follows immediately by adding
Equations~\eqref{eq:pgdot}--\eqref{eq:pedot}.

It is useful to define the total upward and downward rates on the
$i$--$e$ transition,
\begin{equation}
c=c_c+c_\ell,\qquad d=d_c+d_\ell.
\label{eq:cd}
\end{equation}
At stationarity, together with $p_g+p_i+p_e=1$, the~linear equations give
\begin{align}
p_g&=\frac{bc+w(b+d)}{Z},
\label{eq:pgss}\\
p_i&=\frac{ad+w(b+d)}{Z},
\label{eq:piss}\\
p_e&=\frac{ac+w(a+c)}{Z},
\label{eq:pess}
\end{align}
where

\begin{equation}
Z=ac+ad+bc+w(a+c+2b+2d).
\label{eq:Z}
\end{equation}

We next identify the stationary current associated with the useful thermodynamic~cycle.

The probability current through the work transition is
\begin{equation}
I=w(p_i-p_g).
\label{eq:I_def}
\end{equation}
Substituting Equations~\eqref{eq:pgss} and \eqref{eq:piss} gives
\begin{equation}
I=
\frac{w\left[a(d_c+d_\ell)-b(c_c+c_\ell)\right]}{Z}.
\label{eq:I}
\end{equation}
At stationarity the same net useful-cycle current flows through the intended
hot transition.  The~operational output power is therefore defined as
\begin{equation}
P=\epsilon I.
\label{eq:power}
\end{equation}
Positive-power engine operation requires $I>0$.
The direction of the current carried by the parasitic hot transition can be determined analytically from the same stationary~solution.

The net probability current produced by the hot reservoir on the
$i$--$e$ transition is
\begin{equation}
I_\ell=c_\ell p_i-d_\ell p_e.
\label{eq:Ileak_def}
\end{equation}
For thermal reservoirs, detailed balance gives the ratios
\begin{equation}
r_h=e^{-\beta_h\omega_h},\qquad
r_c=e^{-\beta_c\omega_c},\qquad
r_\ell=e^{-\beta_h\omega_c},
\label{eq:rdefs}
\end{equation}
with
\begin{equation}
a=br_h,\qquad
c_c=d_c r_c,\qquad
c_\ell=d_\ell r_\ell.
\label{eq:detailed_balance}
\end{equation}
Using the stationary populations, Equation~\eqref{eq:Ileak_def} becomes
\begin{equation}
I_\ell=
\frac{d_\ell}{Z}
\left[
bd_c r_h(r_\ell-r_c)
+bw(r_\ell-r_h)
+d_cw(r_\ell-r_c)
\right].
\label{eq:Ileak}
\end{equation}

The sign of this expression is fixed under ordinary engine conditions.  Since
$T_h>T_c$ at the same frequency $\omega_c$, one has
$r_\ell>r_c$.  Since $\omega_h>\omega_c$ at the same hot temperature, one also
has $r_\ell>r_h$.  All prefactors in Equation~\eqref{eq:Ileak} are nonnegative, and~therefore
\begin{equation}
I_\ell>0.
\label{eq:Ileak_positive}
\end{equation}
With the sign convention of Equation~\eqref{eq:Ileak_def}, the~parasitic hot
current is always directed from the hot reservoir into the
$i$--$e$ transition.  In~the stationary network that injected energy is
subsequently rejected through the cold channel.  The~unwanted transition is
therefore a genuine hot-to-cold short~circuit.

The corresponding energy currents are
\begin{align}
J_h&=\omega_h I+\omega_c I_\ell,
\label{eq:Jh_markov}\\
J_c&=-\omega_c(I+I_\ell),
\label{eq:Jc_markov}
\end{align}
and satisfy

\begin{equation}
J_h+J_c-P=0,
\label{eq:firstlaw_markov}
\end{equation}
which is the stationary first law of thermodynamics, as~the energy
absorbed from the hot reservoir is partitioned between the heat
released to the cold reservoir and the extracted power.
The efficiency in the positive-power regime is given by
\begin{equation}
\eta=\frac{P}{J_h}
=
\frac{\eta_0}
{1+(\omega_c/\omega_h)(I_\ell/I)}.
\label{eq:eta}
\end{equation}
Equation~\eqref{eq:Ileak_positive} together with Equation (\ref{eq:eta}) shows directly that a nonzero parasitic
current lowers the efficiency below the tight-coupling value
$\eta_0=\epsilon/\omega_h$.

The expression for the useful current also gives a closed condition for the loss of positive-power~operation.

From Equation~\eqref{eq:I}, the~condition $I>0$ is
\begin{equation}
a(d_c+d_\ell)>b(c_c+c_\ell).
\end{equation}
Using Equation~\eqref{eq:detailed_balance}, this reduces to
\begin{equation}
d_c(r_h-r_c)>d_\ell(r_\ell-r_h).
\label{eq:threshold_intermediate}
\end{equation}
For bosonic reservoirs we write the downward rates as
\begin{equation}
b=\frac{\Gamma_h}{1-r_h},\qquad
d_c=\frac{\Gamma_c}{1-r_c},\qquad
d_\ell=\frac{\Gamma_\ell}{1-r_\ell},
\label{eq:bosonic_down}
\end{equation}
where $\Gamma_h$, $\Gamma_c$, and~$\Gamma_\ell$ are the corresponding zero-temperature
rate scales.  The~critical parasitic scale is then

\begin{equation}
\Gamma_\ell^{\rm crit}
=
\Gamma_c
\frac{(r_h-r_c)(1-r_\ell)}
{(1-r_c)(r_\ell-r_h)}.
\label{eq:Gamma_crit}
\end{equation}
When $\Gamma_\ell>\Gamma_\ell^{\rm crit}$, the~useful current changes~sign.

In the absence of the parasitic channel, Equation~\eqref{eq:threshold_intermediate}
reduces to $r_h>r_c$, which is equivalent to

\begin{equation}
\frac{T_h}{T_c}>\frac{\omega_h}{\omega_c}.
\label{eq:engine_condition}
\end{equation}
This is the usual positive-power condition for the ideal three-level engine.
For the finite-bandwidth performance analysis presented in
Section~\ref{sec:results}, we choose a representative parameter set
well inside this engine regime and use $\omega_h$ as the energy scale, with~\begin{equation}
\omega_h=1,\qquad
\epsilon=0.4,\qquad
\omega_c=0.6,
\label{eq:freq_parameters}
\end{equation}
and
\begin{align}
T_h&=0.5,\qquad T_c=0.2,\nonumber\\
\Gamma_h&=\Gamma_c=0.01,\qquad \Gamma_w=0.001.
\label{eq:rate_parameters}
\end{align}
These values satisfy $T_h/T_c=2.5>\omega_h/\omega_c=5/3$ and therefore
correspond to positive-power operation in the absence of the parasitic
transition. For~this reference set, the~critical parasitic rate scale is

\begin{equation}
\Gamma_\ell^{\rm crit}=3.7932\times10^{-3},
\label{eq:crit_numeric}
\end{equation}
or approximately $0.38\,\Gamma_h$. Thus, a parasitic channel substantially
weaker than the intended hot coupling can already alter the thermodynamic
operation of the~engine.

\section{Explicit Finite-Bandwidth~Filter}\label{sec:filter}

{To implement frequency filtering dynamically, we replace the
direct hot reservoir by an auxiliary harmonic mode centered at the
useful transition frequency $\omega_h$. The~enlarged Hamiltonian is
\begin{align}
H ={}&
\omega_h |e\rangle\langle e|
+\epsilon |i\rangle\langle i|
+\omega_h a^\dagger a
\nonumber\\
&+g_h
\left(
|e\rangle\langle g|\,a
+|g\rangle\langle e|\,a^\dagger
\right)
\nonumber\\
&+g_\ell
\left(
|e\rangle\langle i|\,a
+|i\rangle\langle e|\,a^\dagger
\right).
\label{eq:Hfilter}
\end{align}
Here, $a$ is the annihilation operator of the auxiliary mode.
The coupling $g_h$ is the microscopic qutrit--filter coupling
associated with the useful $|g\rangle\leftrightarrow|e\rangle$
transition, which is resonant with the auxiliary mode. The~coupling
$g_\ell$ is the corresponding microscopic coupling of the same
mode to the parasitic $|i\rangle\leftrightarrow|e\rangle$
transition. The~latter transition is detuned from the filter
resonance by

\begin{equation}
\omega_h-\omega_c=\epsilon.
\label{eq:detuning}
\end{equation}

We parameterize the relative strength of the parasitic coupling by
the dimensionless amplitude ratio

\begin{equation}
\chi\equiv \frac{g_\ell}{g_h},
\label{eq:chi}
\end{equation}
so that

\begin{equation}
g_\ell=\chi g_h.
\label{eq:gl}
\end{equation}
Thus, $\chi=1$ corresponds to equal microscopic coupling amplitudes
on the useful and parasitic transitions. This does not imply equal
effective transition rates, because~the two transitions occur at
different frequencies relative to the filter~resonance.

The auxiliary mode loses energy to a residual hot reservoir with
linewidth $\kappa$. If~$g_h$ were held fixed while $\kappa$ were
varied, changing the linewidth would also change the resonant
weak-coupling rate. To~isolate the effect of spectral width, we
therefore impose the constant-Purcell prescription
\clearpage
~
\begin{equation}
g_h=\frac{1}{2}\sqrt{\Gamma_h\kappa},
\label{eq:ghpurcell}
\end{equation}
which keeps
\begin{equation}
\frac{4g_h^2}{\kappa}=\Gamma_h
\label{eq:purcell}
\end{equation}
fixed throughout the bandwidth scan. For~fixed $\chi$, the~parasitic microscopic coupling then scales proportionally,

\begin{equation}
g_\ell
=
\frac{\chi}{2}\sqrt{\Gamma_h\kappa}.
\label{eq:glpurcell}
\end{equation}
Thus, $\Gamma_h$ sets the intended resonant weak-coupling scale,
while $\chi$ fixes the ratio of the    two microscopic qutrit--filter
coupling amplitudes.}

The frequency selectivity and finite response time of the auxiliary
mode follow directly from its Langevin equation.  For~clarity, consider first the useful transition operator    $A_{h}\equiv |g\rangle \langle e|$.  The~mode obeys
\begin{equation}
\dot a(t)=
-\left(\frac{\kappa}{2}+i\omega_h\right)a(t)
-ig_{h}A_{h}(t)
+\sqrt{\kappa}\,a_{\rm in}(t),
\label{eq:langevin}
\end{equation}
where $a_{\rm in}(t)$ is the input noise of the residual hot bath.  The~formal
solution is
\begin{align}
a(t)={}&
e^{-(\kappa/2+i\omega_h)t}a(0)
\nonumber\\
&-ig_{h}\int_0^t ds\,
e^{-(\kappa/2+i\omega_h)(t-s)}A_{h}(s)
\nonumber\\
&+\sqrt{\kappa}\int_0^t ds\,
e^{-(\kappa/2+i\omega_h)(t-s)}a_{\rm in}(s).
\label{eq:langevin_solution}
\end{align}

{ For a Markovian thermal input at temperature $T_h$, the~nonvanishing noise    correlations are
\begin{align}
\langle a_{\rm in}^\dagger(t)a_{\rm in}(t')\rangle
&=n_h\,\delta(t-t'),\\
\langle a_{\rm in}(t)a_{\rm in}^\dagger(t')\rangle
&=(n_h+1)\,\delta(t-t'),
\end{align}
where

\begin{equation}
n_h=\frac{1}{e^{\omega_h/T_h}-1},
\end{equation}
and $\langle a_{\rm in}(t)a_{\rm in}(t')\rangle = \langle a_{\rm in}^\dagger(t)a_{\rm in}^\dagger(t')\rangle=0$. These correlations are the Langevin counterpart of the
thermal auxiliary-mode dissipator introduced in Equation~(53).}

The retarded response is therefore proportional to
\begin{equation}
\chi_a(\tau)=
\Theta(\tau)e^{-\kappa\tau/2}e^{-i\omega_h\tau},
\label{eq:response_time}
\end{equation}
so the characteristic filter response time is of order $2/\kappa$, and~$\Theta(\tau)$ is the step function. Its Fourier transform is given by
\begin{equation}
\widetilde{\chi}_a(\omega)
=
\frac{1}{\kappa/2-i(\omega-\omega_h)},
\label{eq:susceptibility}
\end{equation}
and the corresponding dissipative response is
\begin{equation}
J_{\rm eff}(\omega)
=
g_{h}^2\kappa
\left|\widetilde{\chi}_a(\omega)\right|^2
=
\frac{g_{h}^2\kappa}
{(\kappa/2)^2+(\omega-\omega_h)^2}.
\label{eq:Lorentzian}
\end{equation}
At resonance, Equations~\eqref{eq:purcell} and \eqref{eq:Lorentzian} give
\begin{equation}
J_{\rm eff}(\omega_h)=\Gamma_h.
\end{equation}
At the parasitic frequency $\omega_c=\omega_h-\epsilon$,
\begin{equation}
\frac{J_{\rm eff}(\omega_c)}{J_{\rm eff}(\omega_h)}
=
\frac{\kappa^2}{\kappa^2+4\epsilon^2}.
\label{eq:suppression_factor}
\end{equation}
Equation~\eqref{eq:suppression_factor} is useful for interpreting the
numerical results.  It quantifies the purely spectral benefit of reducing the
linewidth.

For the local-GKSL maximum-power point reported in
Equation~\eqref{eq:gksl_opt}, $\kappa_\star=0.136443$,
Equation~\eqref{eq:suppression_factor} gives

\begin{equation}
\frac{J_{\rm eff}(\omega_c)}{J_{\rm eff}(\omega_h)}
=2.83\times10^{-2}.
\label{eq:suppression_numeric}
\end{equation}
Thus, the filter suppresses the detuned spectral weight to about $2.8\%$ of
its resonant value at the maximum-power~point.

The same parameter $\kappa$ controls both spectral width and
dynamical response. In~the narrow-bandwidth limit, Equation~\eqref{eq:ghpurcell} gives
$g_h\propto\sqrt{\kappa}$, while the residual hot bath dissipator
introduced in Section~\ref{sec:gksl} is proportional to $\kappa$.  The~auxiliary mode therefore becomes weakly
replenished by the hot reservoir and weakly coupled to the qutrit.  The~stationary hot current and output power consequently vanish as the filter is
made sufficiently~narrow.

At larger $\kappa$, the~filter exchanges energy more rapidly with the hot
reservoir.  However, the~suppression factor in
Equation~\eqref{eq:suppression_factor} increases, so the parasitic transition is
progressively reactivated.  The~power is therefore expected to increase from
zero at small bandwidth, reach a maximum at an intermediate linewidth, and~eventually decrease as spectral selectivity is lost.  The~explicit filter results presented in Section~\ref{sec:results}
test this expectation without eliminating the auxiliary~mode.

\section{Open-System Descriptions and Effective~Reduction}
\label{sec:methods}

We now consider three complementary open-system descriptions.
The local-GKSL and nonsecular Redfield approaches retain the enlarged qutrit--filter Hamiltonian $H$, while the effective Lorentzian-rate model is formulated for the reduced three-level system with Hamiltonian $H_A$. All three descriptions use the same reservoir temperatures, coupling ratio $\chi$, and~constant-Purcell protocol; they differ in how the filter dynamics and residual environments are incorporated into the stationary~description.

\subsection{Local GKSL Equation: Primary~Model}
\label{sec:gksl}

Our primary model treats the residual hot bath, cold bath, and~zero-affinity
terminal as weakly coupled Markovian environments acting on the physical filter
and bare qutrit transitions.  The~enlarged density operator obeys
\begin{equation}
\dot\rho=-i[H,\rho]+\mathcal L_h\rho+\mathcal L_c\rho+\mathcal L_w\rho,
\label{eq:master}
\end{equation}
with

\begin{equation}
\D[L]\rho=L\rho L^\dagger-\frac12\{L^\dagger L,\rho\}.
\label{eq:dissipator}
\end{equation}
The residual hot bath acts on the auxiliary mode,
\begin{equation}
\mathcal{L}_h\rho =
\kappa(n_h+1)\mathcal{D}[a]\rho
+\kappa n_h\mathcal{D}[a^\dagger]\rho,
\end{equation}
where $n_h$ is given in Equation~(44).
The cold bath acts on the bare $e$--$i$ transition,
\begin{align}
\mathcal L_c\rho={}\Gamma_c(n_c+1)\D[\ket i\bra e]\rho+\Gamma_c n_c\D[\ket e\bra i]\rho,
\label{eq:Lc}
\end{align}
with $n_c=(e^{\omega_c/T_c}-1)^{-1}$, and~the zero-affinity terminal is
\begin{equation}
\mathcal L_w\rho=\Gamma_w\D[\ket g\bra i]\rho+
\Gamma_w\D[\ket i\bra g]\rho.
\label{eq:Lw}
\end{equation}
This local construction has a direct physical interpretation and is of GKSL
form, so positivity is preserved by construction.  Its quantitative scope is
that of a weak-coupling local master equation: the residual-bath rates must
remain small compared with the bare transition frequencies, while the coherent
couplings $g_{h}$ and $g_{\ell}$ are retained exactly inside $H$.

\subsection{Nonsecular Bloch--Redfield Equation: Independent Dressed-Basis~Check}
\label{sec:redfield}

To test whether the finite-bandwidth optimum depends qualitatively on the local
dissipative construction, we independently treat the same enlarged Hamiltonian
with a frequency-resolved nonsecular Bloch--Redfield equation.  Let
\begin{equation}
H\ket n=E_n\ket n
\end{equation}
and define the Hermitian coupling operators
\begin{equation}
S_h=a+a^\dagger,\;
S_c=\ket e\bra i+\ket i\bra e,\;
S_w=\ket i\bra g+\ket g\bra i.
\end{equation}
For terminal $\nu$ we introduce
\begin{equation}
G_\nu=\sum_{m,n}\gamma_\nu(E_n-E_m)
\langle m|S_\nu|n\rangle\ket m\bra n,
\label{eq:Gnu}
\end{equation}
and use
\begin{align}
\mathcal R_\nu[\rho]=\frac12\big(&G_\nu\rho S_\nu+S_\nu\rho G_\nu^\dagger
-S_\nu G_\nu\rho-\rho G_\nu^\dagger S_\nu\big).
\label{eq:redfield}
\end{align}
For the residual hot and cold baths the rates obey thermal detailed balance,
\begin{equation}
\gamma_\nu(\omega)=
\begin{cases}
\Gamma_\nu[n_\nu(\omega)+1],&\omega>0,\\
\Gamma_\nu n_\nu(|\omega|),&\omega<0,
\end{cases}
\label{eq:redfield_rates}
\end{equation}
with $\Gamma_h$ replaced by $\kappa$ for the residual hot bath.  The~work
terminal is assigned a flat zero-affinity rate $\gamma_w(\omega)=\Gamma_w$ on
the energy-exchanging dressed transitions.  Equation~\eqref{eq:redfield}    retains
cross terms between different dressed Bohr frequencies and is therefore not
   fully~secularized.

The Redfield generator is not GKSL in general.  We therefore use it only as an
independent stationary robustness check and explicitly verify positivity, first-law energy balance,
and second-law Clausius consistency at the representative optima
reported in Section~\ref{sec:thermo_checks}.  The~advantage is that the residual baths are evaluated at the actual dressed
frequencies while interference between nearly degenerate transitions is not
removed by~hand.

\subsection{Effective Lorentzian-Rate Model and Fully Secular~Control}
\label{sec:ratecontrol}

{ The relation between the effective-rate description and the
exact three-state model of Section~\ref{sec:markov} can be made explicit through
a frequency-resolved weak-coupling reduction of the structured
hot environment. The~Lorentzian rate scale follows from
eliminating the auxiliary mode at the level of its susceptibility. Using the filter
susceptibility derived in Section~\ref{sec:filter}, Equation~(46), the~resonant
$|g\rangle\leftrightarrow|e\rangle$ transition acquires the
effective hot    rate scale

\begin{equation}
\Gamma_h^{\rm eff}
=
\frac{4g_{h}^2}{\kappa}
=
\Gamma_h,
\label{eq:Gamma_h_eff}
\end{equation}
where the last equality follows from the constant-Purcell
prescription of Equation~(38).

For the parasitic $|i\rangle\leftrightarrow|e\rangle$ transition,
whose detuning from the filter resonance is
$\omega_c-\omega_h=-\epsilon$, the~corresponding effective rate is
\begin{equation}
\Gamma_\ell(\kappa)
=
\frac{ g_{\ell}^2\kappa}
{(\kappa/2)^2+\epsilon^2}=\frac{ \chi^2g_{h}^2\kappa}
{(\kappa/2)^2+\epsilon^2}
=
\chi^2\Gamma_h
\frac{\kappa^2}
{\kappa^2+4\epsilon^2}.
\label{eq:Gamma_leff}
\end{equation}
Thus, the Lorentzian suppression obtained in Section~\ref{sec:filter} translates,
after adiabatic elimination, into~a bandwidth-dependent parasitic
transition~rate.

The same complex filter response also generates a dispersive
correction. With~the present detuning convention,
\begin{equation}
\Delta_\ell(\kappa)
=-\frac{ g_{\ell}^2\epsilon}
{(\kappa/2)^2+\epsilon^2}=
-\frac{\chi^2 g_{h}^2\epsilon}
{(\kappa/2)^2+\epsilon^2}
=
-\chi^2\Gamma_h
\frac{\kappa\epsilon}
{\kappa^2+4\epsilon^2}.
\label{eq:Delta_leff}
\end{equation}
This term may be interpreted as a filter-induced shift of the
atomic transition. As~in the other stationary calculations of the
present work, we neglect this Hamiltonian renormalization and
retain the dissipative contribution in
Equation~\eqref{eq:Gamma_leff}.

To connect the eliminated filter directly with the thermal rate
model of Section~\ref{sec:markov}, we now use a frequency-resolved adiabatic
description in which the thermal occupation of the hot reservoir
is evaluated at the frequency of each atomic transition. The~resulting reduced master equation for the three-level system is
\begin{align}
\dot{\rho}_A
={}&
-i[H_A,\rho_A]
\nonumber\\
&+
\Gamma_h\big[n_h(\omega_h)+1\big]
{\cal D}[\sigma_{ge}]\rho_A
+
\Gamma_h n_h(\omega_h)
{\cal D}[\sigma_{eg}]\rho_A
\nonumber\\
&+
\Gamma_\ell(\kappa)
\big[n_h(\omega_c)+1\big]
{\cal D}[\sigma_{ie}]\rho_A
+
\Gamma_\ell(\kappa)n_h(\omega_c)
{\cal D}[\sigma_{ei}]\rho_A
\nonumber\\
&+
\Gamma_c\big[n_c(\omega_c)+1\big]
{\cal D}[\sigma_{ie}]\rho_A
+
\Gamma_c n_c(\omega_c)
{\cal D}[\sigma_{ei}]\rho_A
\nonumber\\
&+
\Gamma_w{\cal D}[\sigma_{gi}]\rho_A
+
\Gamma_w{\cal D}[\sigma_{ig}]\rho_A ,
\label{eq:reduced_ME}
\end{align}
where

\begin{equation}
H_A
=
\omega_h|e\rangle\langle e|
+
\epsilon|i\rangle\langle i|,
\end{equation}
$\sigma_{\alpha\beta}=|\alpha\rangle\langle\beta|$, and

\begin{equation}
n_\nu(\omega)
=
\frac{1}{e^{\omega/T_\nu}-1}.
\end{equation}

Equation~\eqref{eq:reduced_ME} should be understood as the
frequency-resolved adiabatic reduction of the structured hot
environment. It is distinct from a direct elimination of the local
dissipator in Equation~(53), for~which the auxiliary-mode thermal
occupation is fixed at $\omega_h$. The~frequency-resolved form is
the appropriate reduced description for establishing the
connection with the thermal detailed-balance structure of Section~\ref{sec:markov}.

The rates appearing in the population equations can then be
identified directly with those introduced in Section~\ref{sec:markov}:
\clearpage
~
\begin{align}
a &=
\Gamma_h n_h(\omega_h),
&
b &=
\Gamma_h\big[n_h(\omega_h)+1\big],
\label{eq:ab_eff}
\\
c_c &=
\Gamma_c n_c(\omega_c),
&
d_c &=
\Gamma_c\big[n_c(\omega_c)+1\big],
\label{eq:cd_c_eff}
\\
c_\ell(\kappa) &=
\Gamma_\ell(\kappa)n_h(\omega_c),
&
d_\ell(\kappa) &=
\Gamma_\ell(\kappa)
\big[n_h(\omega_c)+1\big],
\label{eq:cd_l_eff}
\\
w&=\Gamma_w .
\end{align}
They satisfy

\begin{equation}
\frac{a}{b}=e^{-\beta_h\omega_h},
\qquad
\frac{c_c}{d_c}=e^{-\beta_c\omega_c},
\qquad
\frac{c_\ell}{d_\ell}=e^{-\beta_h\omega_c},
\label{eq:effective_DB}
\end{equation}
which are precisely the detailed-balance relations used in
Equation~(18). Consequently, the~population sector of the
frequency-resolved adiabatic master equation reduces exactly to
Equations~(5)--(7), with~the bandwidth dependence entering through
$\Gamma_\ell(\kappa)$. The~stationary populations, currents,
efficiency, and~engine-reversal condition obtained analytically
in Section~\ref{sec:markov} can therefore be used directly for the effective
Lorentzian-rate~model.

This correspondence gives two additional analytical consequences.
First, let
\begin{equation}
x(\kappa)
\equiv d_\ell(\kappa)
=
\frac{\Gamma_\ell(\kappa)}{1-r_\ell},
\end{equation}
and define

\begin{equation}
A=d_c(r_h-r_c),
\qquad
B=r_\ell-r_h.
\end{equation}
In the engine regime considered here, $A>0$ and $B>0$.
Equation~(14) can then be    written as

\begin{equation}
I(\kappa)
=
\frac{wb\,[A-Bx(\kappa)]}
{Z_0+Z_1x(\kappa)},
\label{eq:I_adiabatic}
\end{equation}
where
\begin{align}
Z_0={}&
a d_c r_c+a d_c+b d_c r_c
+w\big(a+d_c r_c+2b+2d_c\big),
\\
Z_1={}&
a(1+r_\ell)+b r_\ell+w(2+r_\ell).
\end{align}
Since $Z_0>0$ and $Z_1>0$,

\begin{equation}
\frac{\partial I}{\partial x}
=
-
\frac{wb\left(BZ_0+AZ_1\right)}
{\left(Z_0+Z_1x\right)^2}
<0.
\label{eq:dIdx}
\end{equation}
Moreover,

\begin{equation}
\frac{d\Gamma_\ell}{d\kappa}
=
\frac{
8\chi^2\Gamma_h\epsilon^2\kappa
}{
\left(\kappa^2+4\epsilon^2\right)^2
}
>0.
\label{eq:dGammadk}
\end{equation}
It follows that, within~the adiabatically eliminated rate model,
\begin{equation}
\frac{dP_{\rm rate}}{d\kappa}<0
\label{eq:dPdk_rate}
\end{equation}
throughout the positive-power regime. Hence, the reduced
rate model cannot possess an interior maximum of the stationary
power as a function of $\kappa$. Formally, narrowing the filter
always improves the rate-model performance by suppressing the
parasitic channel. The~turnover obtained when the auxiliary mode
is retained explicitly therefore originates from the dynamical
response of the physical filter, which is absent after    adiabatic
elimination.

Second, the~engine-reversal threshold derived in Section~\ref{sec:markov} can be
converted into an analytical critical bandwidth. Setting
\clearpage
~
\begin{equation}
\Gamma_\ell(\kappa_{\rm rev})
=
\Gamma_\ell^{\rm crit}
\end{equation}
and using Equation~\eqref{eq:Gamma_leff} gives
\begin{equation}
\kappa_{\rm rev}
=
2\epsilon
\sqrt{
\frac{\Gamma_\ell^{\rm crit}}
{\chi^2\Gamma_h-\Gamma_\ell^{\rm crit}}
},
\label{eq:kappa_rev}
\end{equation}
provided
$\chi^2\Gamma_h>\Gamma_\ell^{\rm crit}$.
If
$\chi^2\Gamma_h\leq\Gamma_\ell^{\rm crit}$,
the effective parasitic rate remains below the reversal threshold
for all~bandwidths.

The adiabatic description should, however, be interpreted within
its regime of validity. Under~the constant-Purcell prescription,
\begin{equation}
\frac{g_{h}}{\kappa}
=
\frac{1}{2}
\sqrt{\frac{\Gamma_h}{\kappa}},
\end{equation}
which increases as the filter is narrowed. Consequently,
sufficiently small $\kappa$ eventually lies outside the regime in
which the auxiliary mode can be regarded as instantaneously
following the qutrit. The~formal boundary optimum
$\kappa\rightarrow0$ of the rate model should therefore not be
interpreted as a controlled prediction of the adiabatic
approximation in that limit. It instead emphasizes precisely why
the explicit dynamical treatment of the auxiliary mode is required
to describe the narrow-bandwidth regime.}

For completeness, we also tested a fully secular global
generator built from the dressed transitions of $H$, whose
construction is given explicitly in Appendix~\ref{app:secular}.
It is completely positive, but~in the present near-degenerate
weak-coupling limit it does not recover the elementary no-leak
three-state engine. We therefore show it only in the weak-coupling validation test and do not use it for quantitative performance~predictions.

\subsection{Currents, Power, and~Stationary~Consistency}

For the local GKSL description the stationary state satisfies
\begin{equation}
0=-i[H,\rho_{\rm ss}]+\sum_\nu\mathcal L_\nu\rho_{\rm ss},
\end{equation}
and the terminal currents are

\begin{equation}
J_\nu=\Tr\!\left[H\,\mathcal L_\nu(\rho_{\rm ss})\right].
\label{eq:currents}
\end{equation}
For Redfield, we replace $\mathcal L_\nu$ by $\mathcal R_\nu$ in the same
definition.  With~$J_\nu>0$ denoting energy supplied to the enlarged system,
\begin{equation}
P=-J_w,\qquad \eta=\frac{P}{J_h},\qquad J_h>0,\ P>0.
\label{eq:performance}
\end{equation}
At stationarity, the~terminal currents must satisfy the first-law
energy balance
\begin{equation}
J_h+J_c+J_w=0,
\label{eq:firstlaw_full}
\end{equation}
which is equivalent to $J_h+J_c-P=0$ because $P=-J_w$.
To assess consistency with the second law of thermodynamics, we also
monitor the stationary Clausius entropy-production quantity

\begin{equation}
\dot\Sigma_C
=
-\frac{J_h}{T_h}
-\frac{J_c}{T_c}.
\label{eq:clausius}
\end{equation}
For thermodynamically consistent operation, one expects
$\dot\Sigma_C\geq0$. For~the local GKSL model, however, this quantity
is used here as a stationary second-law consistency diagnostic rather
than as a general theorem for local master equations; for Redfield, it
is likewise a diagnostic rather than a consequence of complete
positivity.

\section{Numerical Construction and~Validation}
\label{sec:validation}

The oscillator is truncated to the first $N$ Fock states, giving Hilbert-space
dimension $3N$.  For~each $\kappa$ and $\chi$, we set
$g_{h}=\sqrt{\Gamma_h\kappa}/2$, $g_{\ell}=\chi g_{h}$, and~construct $H$, build the chosen dissipative
generator, vectorize the stationary equation, replace one row by
$\Tr\rho_{\rm ss}=1$, and~solve the resulting linear system.  The~local-GKSL
vectorization is given in Appendix~\ref{app:vec}; the Redfield construction is
summarized in Appendix~\ref{app:redfield}.

\begin{figure}[H]
\centering
\includegraphics[width=0.99\textwidth]{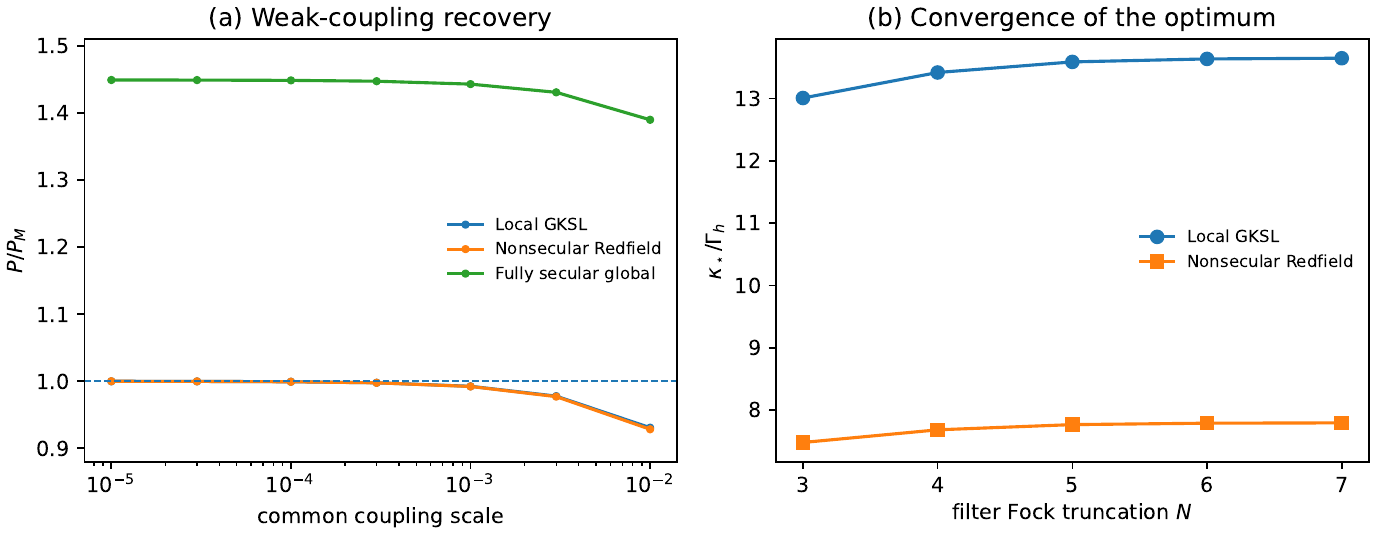}
\caption{Validation
 of the dynamical descriptions.  (\textbf{a}) Recovery of the exact
no-leak Markovian engine as the common coupling scale $\Gamma$ is reduced at
$\chi=0$ and $\kappa=0.1$.  The~local GKSL and nonsecular Redfield treatments
approach $P/P_M=1$ marked by the blue dashed line, whereas the fully secular global construction does not.
(\textbf{b}) Convergence of the maximum-power bandwidth with filter Fock truncation for
$\chi=1$. Both retained explicit filter descriptions converge rapidly,
although to different quantitative optima.}
\label{fig:validation}
\end{figure}

A necessary check is recovery of the elementary no-leak Markovian engine.  We
set $\chi=0$, $\kappa=0.1$ and scale
\begin{equation}
\Gamma_h=\Gamma_c=\Gamma,\qquad \Gamma_w=\Gamma/10
\end{equation}
toward zero.  Figure~\ref{fig:validation}a shows that both the local GKSL and
nonsecular Redfield results approach the exact Markovian power $P_M$ and reach
$P/P_M=0.999850$ and $0.999847$, respectively, at~$\Gamma=10^{-5}$.  By~contrast, the~fully secular global control approaches approximately $1.449$
rather than unity.  This failure is the reason it is not retained as a
quantitative model~here.

Figure~\ref{fig:validation}b shows the independent
Fock-space convergence of the two retained explicit filter
descriptions.  From~$N=6$ to $N=7$, the~local-GKSL optimum changes
from $13.6332$ to $13.6443$ in units of $\Gamma_h$, while the Redfield optimum
changes from $7.7887$ to $7.7938$.  At~the local-GKSL optimum the mean filter
occupation is $0.15541$ and the highest retained state $n=6$ carries only
$5.14\times10^{-6}$ probability.

\section{Finite-Bandwidth Protection: Comparison of~Methods}
\label{sec:results}

Unless otherwise stated we use
$\omega_h=1$, $\epsilon=0.4$, $\omega_c=0.6$, $T_h=0.5$, $T_c=0.2$,
$\Gamma_h=\Gamma_c=0.01$, $\Gamma_w=0.001$, and~$N=7$.

\subsection{Bandwidth~Dependence}
\begin{figure}[H]
\centering
\includegraphics[width=0.94\textwidth]{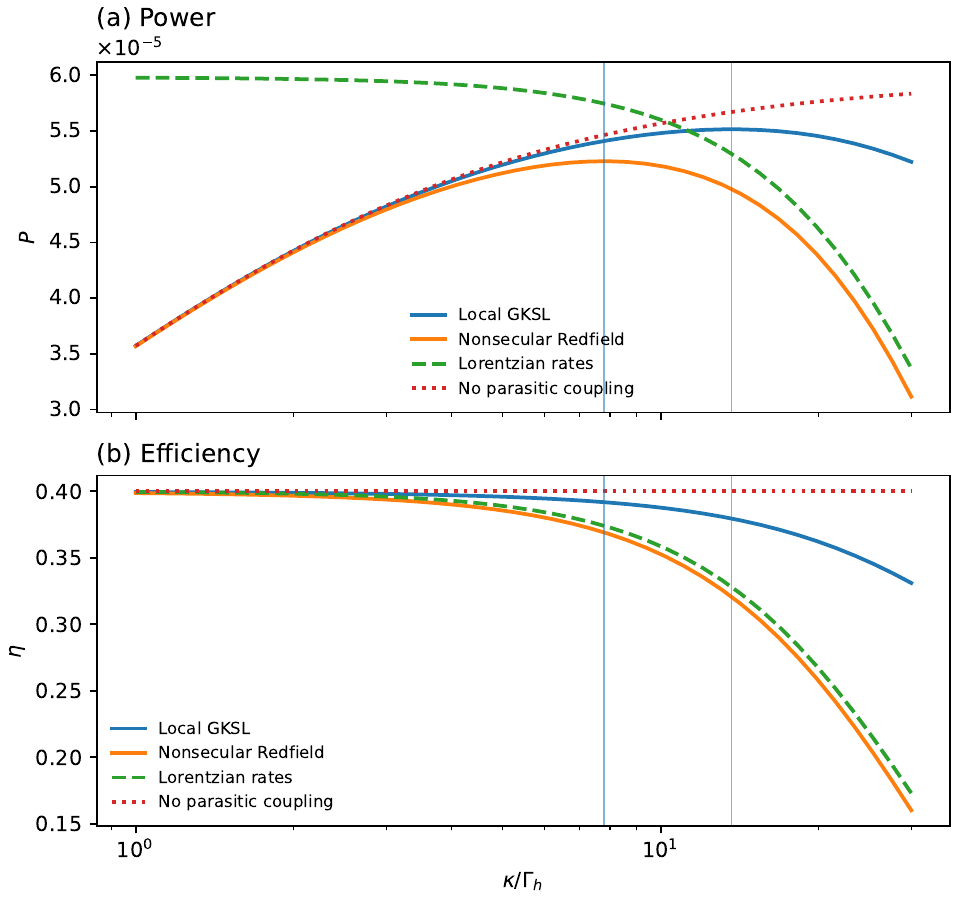}
\caption{Method
 comparison for $\chi=1$.  (\textbf{a}) Stationary power and (\textbf{b})
efficiency versus filter bandwidth.  The~local GKSL equation is the primary
explicit filter model; the nonsecular Redfield curve is an independent
frequency-resolved check; the Lorentzian-rate curve is the frequency-resolved
effective-rate reduction; and the no-parasitic curve is the local-GKSL reference at the same
bandwidth.  The~two thin vertical markers indicate the Redfield and GKSL
maximum-power bandwidths, respectively.  The~displayed range is restricted to
$\kappa\le0.3$.}
\label{fig:methods}
\end{figure}

Figure~\ref{fig:methods} compares all retained descriptions for $\chi=1$.
The local-GKSL and nonsecular Redfield calculations both exhibit
an interior maximum of the stationary power. By~contrast, the~Lorentzian-rate result decreases monotonically with $\kappa$, as~proved analytically in    Equation~(\ref{eq:dPdk_rate}), and~therefore has no interior
maximum.

The converged local-GKSL maximum is
\begin{align}
\frac{\kappa_\star^{\rm GKSL}}{\Gamma_h}&=13.6443,\nonumber\\
P_\star^{\rm GKSL}&=5.5153\times10^{-5},\qquad
\eta_\star^{\rm GKSL}=0.37950.
\label{eq:gksl_opt}
\end{align}
At the same bandwidth the local-GKSL engine with the parasitic coupling removed
has $P_0=5.6709\times10^{-5}$ and $\eta_0=0.4000$, hence
\begin{equation}
\frac{P_\star^{\rm GKSL}}{P_0}=0.9726,
\qquad
\frac{\eta_\star^{\rm GKSL}}{\eta_0}=0.9488.
\end{equation}
The corresponding spectral suppression is
$J_{\rm eff}(\omega_c)/J_{\rm eff}(\omega_h)=2.83\times10^{-2}$.

The independent Redfield calculation gives
\begin{align}
\frac{\kappa_\star^{\rm R}}{\Gamma_h}&=7.7938,\nonumber\\
P_\star^{\rm R}&=5.2288\times10^{-5},\qquad
\eta_\star^{\rm R}=0.36913.
\label{eq:redfield_opt}
\end{align}
At its own optimum the corresponding no-leak Redfield calculation gives
$P_0^{\rm R}=5.4528\times10^{-5}$    and $\eta_0^{\rm R}=0.39982$, so the
retained fractions are $0.9589$ in power and $0.9232$ in efficiency.  The~numerical optimum therefore shifts appreciably between dissipative
constructions, but~the finite optimum, the~protected high-performance regime,
and the underlying bandwidth competition~survive.

\subsection{Robustness with Parasitic Coupling and Thermal~Bias}

\begin{figure}[H]
\centering
\includegraphics[width=0.99\textwidth]{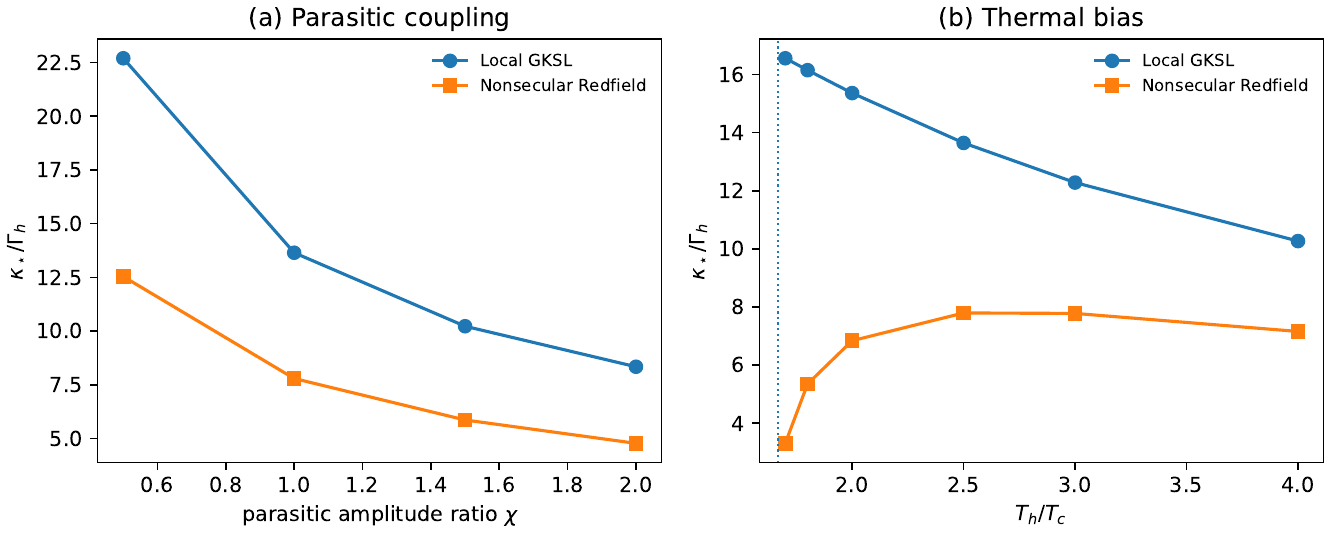}
\caption{Robustness of the optimal-bandwidth trend.  (\textbf{a}) Maximum-power
bandwidth versus parasitic microscopic amplitude ratio $\chi$.  (\textbf{b}) Maximum-power
bandwidth versus thermal bias at fixed $T_c=0.2$ and $\chi=1$.  The~local GKSL
and nonsecular Redfield descriptions differ quantitatively but predict the same
finite-bandwidth protection mechanism and the same shift toward narrower filters
as the parasitic coupling is increased.  The~vertical dotted line in panel (\textbf{b})
marks the no-leak threshold $T_h/T_c=5/3$.}
\label{fig:robustness}
\end{figure}

A useful robustness criterion is whether the two explicit filter descriptions
predict the same systematic movement of the optimum as physical parameters are
changed.     Figure~\ref{fig:robustness}a shows that they do.  As~$\chi$
increases from $0.5$ to $2$, the~GKSL optimum moves from  $\kappa_\star/\Gamma_h=22.69$ to $8.34$, while the Redfield optimum moves from
$12.54$ to $4.78$.  Stronger parasitic coupling therefore consistently favors a
narrower~filter.

The thermal-bias comparison in Figure~\ref{fig:robustness}b is likewise
informative.  In~the primary GKSL model the optimum decreases monotonically from
$16.56$ at $T_h/T_c=1.7$ to $10.27$ at $T_h/T_c=4$.  Redfield predicts a
stronger near-threshold shift but again a finite optimum throughout the engine
regime, with~$\kappa_\star/\Gamma_h=7.79$ at the representative ratio $2.5$.
The exact numerical ridge is therefore method dependent, especially close to
threshold, whereas the existence of a finite protected operating scale is~not. Some comparative values are given in Table~1.

\begin{table}[H]
\caption{Comparison
 of the descriptions for $\chi=1$.  The~rate-only model has
no interior optimum because the explicit filter response time has been
eliminated.  The~fully secular global model is excluded from performance
comparisons because it fails the weak-coupling recovery test of
Figure~\ref{fig:validation}a.}
\label{tab:methods}
\centering
\begin{tabular}{lccc}
\toprule
\textbf{Method} & \boldmath{$\kappa_\star/\Gamma_h$} & \boldmath{$P_\star$} & \boldmath{$\eta_\star$}\\
\midrule
Local GKSL & 13.6443 & $5.5153\times10^{-5}$ & 0.37950\\
Nonsecular Redfield & 7.7938 & $5.2288\times10^{-5}$ & 0.36913\\
Lorentzian rates & $\kappa\to0$ & $5.9824\times10^{-5}$ & 0.4000\\
Fully secular global & excluded & -- & --\\
\bottomrule
\end{tabular}
\end{table}

\subsection{Stationary Thermodynamic~Checks}\label{sec:thermo_checks}
At the local-GKSL optimum the currents are
\clearpage
~
\begin{align}
J_h&=1.45330\times10^{-4},\nonumber\\
J_c&=-9.01772\times10^{-5},\qquad
J_w=-5.51531\times10^{-5}.
\end{align}
Using the unrounded numerical values, the~first-law residual is below
$10^{-17}$, confirming stationary energy conservation, while
$\dot\Sigma_C=1.6023\times10^{-4}>0$, consistent with the
second-law Clausius condition. The~minimum stationary eigenvalue is
$2.37\times10^{-7}$, while positivity is guaranteed by the GKSL~form.

At the Redfield optimum, the~corresponding currents are
\begin{equation}
J_h=1.41650\times10^{-4},\qquad
J_c=-8.93628\times10^{-5},\qquad
J_w=-5.22876\times10^{-5}.
\end{equation}
Using the unrounded numerical values, the~first-law residual is
$2.5\times10^{-17}$, while
$\dot\Sigma_C=1.6351\times10^{-4}>0$, again consistent with the
second-law Clausius condition. The~minimum stationary eigenvalue is
$2.45\times10^{-7}$. Thus the independent Redfield stationary state
is positive in the reported regime despite the absence of a general
complete-   positivity~guarantee.

\section{Discussion}
\label{sec:discussion}
The comparison separates the robust physical statement from the model-dependent
quantitative details.  The~robust statement is that the filter performs two
opposing functions.  Its Lorentzian spectrum suppresses the unwanted transition
at $\omega_c$, while its finite damping and the scaling
$g_{h}\propto\sqrt{\kappa}$ impose a transport cost when the filter is too narrow.
Both explicit filter descriptions therefore generate an interior maximum,
whereas the rate-only reduction cannot because it removes the slow filter degree
of~freedom.

The local GKSL and nonsecular Redfield calculations differ in how the residual
baths resolve the interacting qutrit--filter system.  The~local GKSL equation
couples the baths to the physical filter and bare qutrit transitions and has the
important advantage of complete positivity.  The~Redfield calculation instead
evaluates thermal rates at the dressed Bohr frequencies and retains
near-degenerate cross-frequency terms.  The~shift from
   $\kappa_\star/\Gamma_h=13.64$ to $7.79$ is therefore a useful estimate of the
quantitative sensitivity of the optimum to the dissipative construction, not a
failure of the filtering~mechanism.

The weak-coupling analysis gives additional guidance.  Both retained dynamical descriptions
connect continuously to the exact no-leak three-state engine, while a fully secular
global construction does not.  The~latter result illustrates that complete
positivity alone does not guarantee a uniform approximation when dressed
splittings become nearly degenerate.  Conversely, the~Redfield comparison shows
that the primary GKSL result is not merely an artifact of the local dissipators:
the finite optimum and its systematic parameter trends persist when the baths
are treated in the dressed~basis.

The result should not be described as a generic benefit of
``non-Markovianity.''  The auxiliary mode produces a finite correlation time,
but the relevant mechanism is more concrete: a physical frequency filter is
also an element of the transport network.  A~phenomenological memory kernel or
a stationary Lorentzian rate can reproduce spectral discrimination without
reproducing the same dynamical~bottleneck.

The zero-affinity terminal is likewise an operational energy repository rather
than a microscopic battery.  Replacing it by an explicit work-storage degree of
freedom would sharpen the notion of extractable work~\cite{Niedenzu2019}, but~is
not required to identify the heat leak protection mechanism.  Experimentally,
filtering resonators and engineered electromagnetic environments provide a
natural route to the regime considered here.  The~design rule is direct: the linewidth should resolve
$\epsilon=\omega_h-\omega_c$ sufficiently well to suppress the parasitic transition while remaining large enough to sustain useful hot-side~throughput.

\subsection*{{Possible}
 Superconducting-Circuit~Implementation}
\label{sec:implementation}

The finite-bandwidth protection mechanism considered here is well
matched to present superconducting-circuit reservoir engineering
platforms. Recent experiments provide    two particularly relevant
points of contact. Uusn\"akki~et~al. demonstrated a cyclic
quantum Otto engine using a flux-tunable transmon and a
voltage-controlled quantum-circuit refrigerator (QCR) as an engineered
thermal reservoir~\cite{Uusnakki2026NatCommun}. In a complementary recent experiment, an~autonomous superconducting quantum heat engine was realized using
coupled resonators interacting with spectrally filtered hot and cold
reservoirs~\cite{Uusnakki2026}. These experiments establish
that tunable thermal microwave environments, auxiliary filtering
resonators, and~autonomous heat-to-work conversion are experimentally
accessible in circuit~QED.

The thermodynamic architecture studied here is nevertheless distinct
from both implementations. The~QCR experiment of
Ref.~\cite{Uusnakki2026NatCommun} realizes sequential heating,
cooling, and~flux-control strokes of an Otto cycle. In~the autonomous
device of Ref.~\cite{Uusnakki2026}, distinct filtered hot
and cold reservoirs act on a resonator-based working medium whose
internal dynamics implement an approximate autonomous Otto cycle.
Here, by~contrast, the~working medium is a stationary three-level
system with simultaneously active hot, cold, and~output channels.
Most importantly, a~\emph{single} structured hot environment couples
resonantly to the useful $|g\rangle\leftrightarrow|e\rangle$
transition and parasitically to the
$|i\rangle\leftrightarrow|e\rangle$ transition. The~proposed
experiment therefore targets a different question: whether tuning the
bandwidth of that hot environment can suppress a specific
hot-to-cold thermodynamic shortcut without throttling the useful~cycle.

\begin{figure}[H]
\centering

\centering
\includegraphics[width=0.70\textwidth]{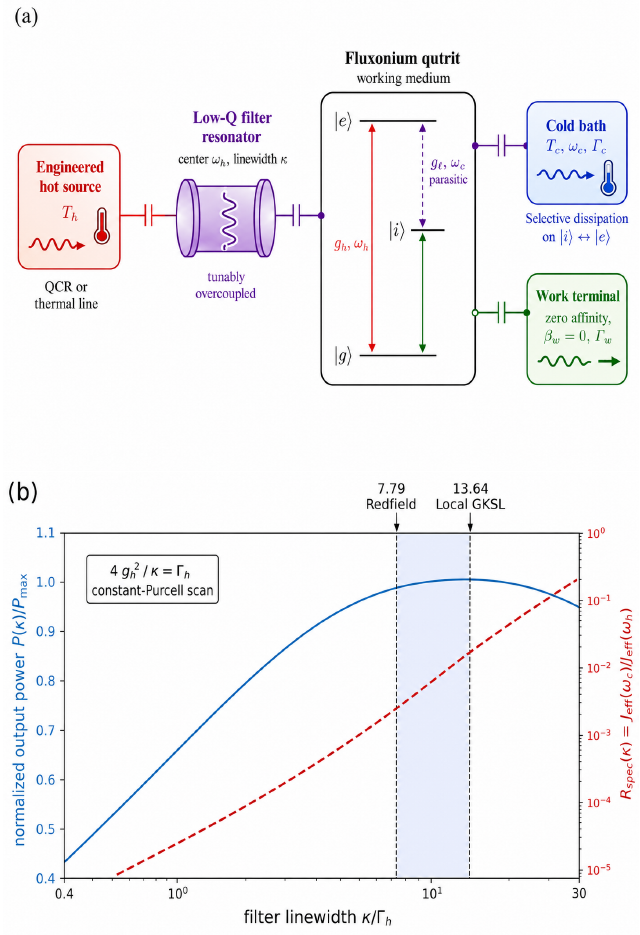}
\caption{Proposed superconducting-circuit test of finite-bandwidth protection.
(\textbf{a}) A strongly anharmonic three-level artificial atom, here represented
by a fluxonium qutrit, is coupled to a tunable low-$Q$ auxiliary resonator
centered at $\omega_h$ with linewidth $\kappa$. The resonator acts as the
structured hot filter and couples resonantly to the useful
$|g\rangle\leftrightarrow|e\rangle$ transition with microscopic amplitude
$g_h$, while coupling off-resonantly to the parasitic
$|i\rangle\leftrightarrow|e\rangle$ transition with microscopic amplitude
$g_\ell$. An engineered thermal microwave source supplies the hot reservoir,
while independent channels provide selective cold dissipation on the
$|i\rangle\leftrightarrow|e\rangle$ transition and the zero-affinity output
terminal on $|g\rangle\leftrightarrow|i\rangle$, with
$\epsilon=\omega_h-\omega_c$.
(\textbf{b}) Experimental signature under a constant-Purcell scan,
$4g_h^2/\kappa=\Gamma_h$. The normalized stationary output power (blue solid
line) develops a maximum at finite linewidth, whereas the residual
spectral-weight ratio
$R_{\rm spec}(\kappa)=J_{\rm eff}(\omega_c)/J_{\rm eff}(\omega_h)
=\kappa^2/(\kappa^2+4\epsilon^2)$ (red dashed line) increases monotonically
with $\kappa$, corresponding to improved spectral rejection as the filter is
narrowed. The shaded interval marks the representative optimum window bounded
by the nonsecular Redfield and local-GKSL predictions,
$\kappa_\star^{\rm R}/\Gamma_h=7.79$ and
$\kappa_\star^{\rm GKSL}/\Gamma_h=13.64$, respectively.}
\label{fig:experimental_implementation}
\end{figure}

A natural implementation would use a strongly anharmonic
superconducting artificial atom operated as an effective qutrit.
Fluxonium is an attractive candidate because its level structure and
selection rules can be engineered to access transition configurations
that are unavailable in a nearly harmonic ladder; in particular,
$\Lambda$-type three-level structures and otherwise forbidden
transitions have been demonstrated experimentally
~\cite{Vool2018}. The~useful transition
$|g\rangle\leftrightarrow|e\rangle$ would be coupled to a deliberately
low-$Q$ auxiliary microwave resonator of frequency $\omega_h$ and
externally controlled linewidth $\kappa$. This resonator plays the
role of the structured hot filter. The~same resonator field also
couples off resonantly to
$|i\rangle\leftrightarrow|e\rangle$, thereby realizing the intended
and parasitic matrix elements $g_{h}$ and $g_{\ell}$ of Equation~\eqref{eq:Hfilter}.
A separate frequency-selective dissipative channel coupled to the
$i$--$e$ transition provides the cold reservoir, while the
$g$--$i$ transition is connected to an output channel implementing
the zero-affinity terminal used in the present stationary model.
Figure~\ref{fig:experimental_implementation} summarizes a possible
superconducting-circuit realization of the finite-bandwidth protection
mechanism and the corresponding experimental~signature.

The experimentally relevant distinction between the rate-only and
explicit filter descriptions can then be tested directly. According
to Equation~(49), the~parasitic spectral weight decreases monotonically
as the filter is narrowed. If~the resonator is retained dynamically,
however, both explicit filter descriptions predict that the output
power eventually decreases for sufficiently small $\kappa$.
The experimentally decisive signature is therefore the simultaneous
observation of improving spectral rejection and a turnover of the
stationary output power at a finite filter~linewidth.

Representative physical scales can be obtained by choosing
\begin{equation}
\omega_h/2\pi = 6~{\rm GHz},
\end{equation}
while retaining the dimensionless parameters used throughout the
calculations. This gives
\begin{equation}
\omega_c/2\pi = 3.6~{\rm GHz},
\qquad
\epsilon/2\pi = 2.4~{\rm GHz},
\end{equation}
and, from~$\Gamma_h/\omega_h=0.01$,
\begin{equation}
\Gamma_h/2\pi \simeq 60~{\rm MHz}.
\end{equation}
The two explicit filter calculations then locate the maximum-power
bandwidth    at approximately
\begin{equation}
\kappa_\star/2\pi \simeq 0.47~{\rm GHz}
\qquad
\text{(nonsecular Redfield)}
\end{equation}
and
\begin{equation}
\kappa_\star/2\pi \simeq 0.82~{\rm GHz}
\qquad
\text{(local GKSL)}.
\end{equation}
Equivalently, the~predicted optimum lies in the representative window
\begin{equation}
7.8
\lesssim
\frac{\kappa_\star}{\Gamma_h}
\lesssim
13.6,
\label{eq:experimental_window}
\end{equation}
corresponding to resonator quality factors of order $Q\simeq7$--$13$.
This interval should not be interpreted as universal; rather, it gives
a realistic target range while also quantifying the sensitivity of the
optimal linewidth to the treatment of the residual~baths.

Under the constant-Purcell protocol of Equation~(38), the~corresponding
coherent coupling strengths are of order
\begin{equation}
g_{h}/2\pi \simeq 84\text{--}111~{\rm MHz}.
\end{equation}
The temperatures $T_h=0.5\,\omega_h$ and $T_c=0.2\,\omega_h$
translate to approximately $144~{\rm mK}$ and $58~{\rm mK}$,
respectively. These numbers are intended as representative circuit-QED
scales rather than as a device-specific~design.

A proof-of-principle experiment would therefore implement the
constant-Purcell protocol of Equation~(38) while sweeping the externally
controlled filter linewidth $\kappa$.
Implementing this protocol would require either a tunable
qutrit--filter coupler or flux control of the relevant transition
matrix element so that $g_h$ can be varied independently of the
filter's external damping. For~each value of
$\kappa$, one would measure the stationary output power together with
the response of the hot channel at $\omega_c$. The~characteristic
signature would be that the parasitic spectral weight continues to
decrease as $\kappa$ is reduced, whereas the power reaches a maximum
within a finite-bandwidth window and subsequently decreases. Such a
measurement would directly distinguish the dynamical throughput cost
of a physical filter from the purely spectral protection captured by
an adiabatically eliminated rate~model.
\section{Conclusions}
\label{sec:conclusion}
We have established an analytically grounded and numerically cross-validated picture of
finite-bandwidth protection in a three-level quantum heat engine.  The~exact
three-state network proves that the unwanted hot transition is a hot-to-cold short
circuit and yields a closed threshold for engine reversal.  An~explicit damped
filter then reveals the complementary cost of implementing spectral selectivity:
an excessively narrow filter becomes dynamically~slow.

The central finding is independent of a single open-system approximation.  A~completely positive local GKSL model and an independent nonsecular Redfield
calculation both recover the exact no-leak Markovian engine in the appropriate
weak-coupling limit and both predict a finite maximum-power bandwidth.  For~equal
useful and parasitic microscopic coupling amplitudes they give, respectively,
$\kappa_\star/\Gamma_h=13.6443$ and $7.7938$.  The~quantitative displacement of
the optimum identifies the sensitivity to bath modeling, while the existence of
the optimum, its high no-leak power retention, and~its systematic shift with
parasitic coupling are~robust.

The resulting reservoir engineering principle is therefore specific: spectral
filtering can protect a continuous quantum heat engine from a parasitic
thermodynamic cycle, but~a real filter cannot be narrowed without a dynamical
cost.  The~useful operating point is set by the balance between spectral
rejection and energy~throughput.

\section*{Author Contributions}
All authors have contributed equally for the development of this work.

\section*{Funding}
M.C.d.O. acknowledges partial financial support from the National Institute of Science and Technology for Applied Quantum Computing through CNPq (Process No.~408884/2024-0) and from the S\~ao Paulo Research Foundation (FAPESP) through the Center for Research and Innovation on Smart and Quantum Materials (CRISQuaM, Process No.~2024/00998-6).



\section*{Data Availability Statement}
The data generated and analyzed during this study are available from the corresponding author upon reasonable request.


\appendix
\setcounter{equation}{0}
\renewcommand{\theequation}{A\arabic{equation}}
\renewcommand{\theHequation}{A\arabic{equation}}

\section{Vectorized Local-GKSL Stationary~Equation}
\label{app:vec}
For

 column vectorization,
\begin{equation}
\operatorname{vec}(A\rho B)=(B^T\otimes A)\operatorname{vec}(\rho).
\end{equation}
The Hamiltonian contribution is
\begin{equation}
\mathcal L_H=-i(I\otimes H-H^T\otimes I),
\end{equation}
and one Lindblad jump operator contributes
\begin{equation}
\mathcal D_L=L^*\otimes L-\frac12 I\otimes L^\dagger L
-\frac12(L^\dagger L)^T\otimes I.
\end{equation}
The full stationary Liouvillian is the sum of the Hamiltonian and all jump
contributions.  One row is replaced by the trace-normalization~condition.

\section{Nonsecular Redfield~Implementation}
\label{app:redfield}
In the eigenbasis of $H$, Equation~\eqref{eq:Gnu} is evaluated element by element at
the actual Bohr frequency $E_n-E_m$.  The~superoperator corresponding to
Equation~\eqref{eq:redfield} is then added to the Hamiltonian Liouvillian without
frequency grouping or full secularization.  The~resulting stationary linear
system is solved with the same trace replacement as in the local-GKSL
calculation.  Hermiticity is symmetrized only at numerical roundoff before
stationary eigenvalues and currents are evaluated.  For~the weak-coupling calculation in
Figure~\ref{fig:validation}a, the~Redfield ratios $P/P_M$ are
$0.928327$, $0.976988$, $0.992123$, $0.997568$, $0.999138$, $0.999689$, and~$0.999847$ as $\Gamma$ is reduced from $10^{-2}$ to $10^{-5}$; the corresponding
local-GKSL values are $0.930703$, $0.977731$, $0.992374$, $0.997644$,
$0.999163$, $0.999697$, and~$0.999850$.

\section{Fully Secular Global~Control}
\label{app:secular}

For completeness, we specify the fully secular global
construction used only as the diagnostic control in
Figure~\ref{fig:validation}.  The~starting point is the same
enlarged qutrit--filter Hamiltonian $H$ of Equation~\eqref{eq:Hfilter},
which is diagonalized as
\begin{equation}
    H |n\rangle = E_n |n\rangle .
    \label{eq:sec_eigen}
\end{equation}
The system coupling operators are the same as those used in
the nonsecular    Redfield calculation,
\begin{equation}
    S_h = a+a^\dagger ,
    \qquad
    S_c = |e\rangle\langle i|+|i\rangle\langle e| ,
    \qquad
    S_w = |i\rangle\langle g|+|g\rangle\langle i| .
    \label{eq:sec_couplings}
\end{equation}

For each terminal $\nu\in\{h,c,w\}$ and each Bohr frequency
$\omega$, we introduce the dressed transition operator

\begin{equation}
    A_\nu(\omega)
    =
    \sum_{E_n-E_m=\omega}
    |m\rangle\langle m|
    S_\nu
    |n\rangle\langle n| .
    \label{eq:sec_jump}
\end{equation}
Thus, for~$\omega>0$, $A_\nu(\omega)$ lowers the energy of
the enlarged system by $\omega$.  Transitions sharing the
same Bohr frequency are retained within the same operator
$A_\nu(\omega)$, while all terms involving distinct Bohr
frequencies are discarded by the full secular~approximation.

The corresponding global GKSL generator is
\begin{equation}
    \mathcal{L}^{\rm sec}_\nu[\rho]
    =
    \sum_{\omega}
    \gamma_\nu(\omega)
    \left[
        A_\nu(\omega)\rho A_\nu^\dagger(\omega)
        -
        \frac{1}{2}
        \left\{
            A_\nu^\dagger(\omega)A_\nu(\omega),
            \rho
        \right\}
    \right] ,
    \label{eq:sec_generator}
\end{equation}
and the stationary state satisfies
\begin{equation}
    0
    =
    -i[H,\rho_{\rm ss}]
    +
    \sum_{\nu=h,c,w}
    \mathcal{L}^{\rm sec}_\nu[\rho_{\rm ss}] .
    \label{eq:sec_stationary}
\end{equation}
No Lamb-shift contribution is included, consistently with the
other stationary comparisons in the~manuscript.

For the residual hot and cold reservoirs, we use the same
frequency-resolved thermal rates as in the Redfield
calculation,
\begin{equation}
\gamma_\nu(\omega)
=
\begin{cases}
\Gamma_\nu
\left[n_\nu(\omega)+1\right],
&
\omega>0,
\\[3pt]
\Gamma_\nu
n_\nu(|\omega|),
&
\omega<0,
\end{cases}
\label{eq:sec_rates}
\end{equation}
where

\begin{equation}
    n_\nu(\omega)
    =
    \frac{1}{e^{\omega/T_\nu}-1}.
\end{equation}
For the residual hot bath one sets $\Gamma_h\rightarrow\kappa$,
as in Equation~\eqref{eq:redfield_rates}.  The~work terminal is
assigned the same flat zero-affinity rate used in the
Redfield calculations,
\begin{equation}
    \gamma_w(\omega)=\Gamma_w
\end{equation}
on the energy-exchanging dressed~transitions.

The essential difference from the nonsecular Redfield
construction is therefore not the bath spectra, temperatures,
or system Hamiltonian, but~the removal of every
cross-frequency contribution with
$\omega\neq\omega'$.  Schematically, the~Redfield generator
contains terms of the form

\begin{equation}
    A_\nu(\omega)\rho A_\nu^\dagger(\omega'),
    \qquad
    \omega\neq\omega',
\end{equation}
whereas Equation~\eqref{eq:sec_generator} retains only
$\omega=\omega'$.  The~secular generator is consequently of
GKSL form and completely~positive.

This construction is normally appropriate when distinct
Bohr frequencies are separated by scales large compared with
the dissipative broadening. However, in~the present weak-coupling limit, some dressed splittings collapse together with the overall coupling scale.  Full secularization then removes
cross-frequency terms before this limiting degeneracy is
resolved.  As~shown in Figure~\ref{fig:validation}a, the~resulting stationary power does not approach the exact no-leak
three-state Markovian value: in the weakest-coupling point
shown,
\begin{equation}
    \frac{P_{\rm sec}}{P_M}\simeq 1.449,
\end{equation}
whereas both the local-GKSL and nonsecular Redfield
descriptions approach unity.  We therefore use the fully
secular construction only as a diagnostic of the
near-degenerate secular approximation and not for the
quantitative performance comparisons of Section~\ref{sec:results}.

\end{document}